\documentstyle[12pt]{article}

\begin{document}

\baselineskip 24pt

\title{Applications of a Simple Formula}
\author{Lior Goldenberg and Lev Vaidman}
\date{}
\maketitle

\begin{center}
{\small \em School of Physics and Astronomy \\
Raymond and Beverly Sackler Faculty of Exact Sciences \\
Tel Aviv University, Tel-Aviv 69978, Israel. \\}
\end{center}

\vspace{2cm}
\begin{abstract}

New applications of the formula $A |\psi\rangle = \langle A \rangle
|\psi\rangle + \Delta A |\psi_{\perp} \rangle$ are discussed. Simple
derivations of the Heisenberg uncertainty principle and of related
inequalities are presented. In addition, the formula is used in an
instructive paradox which clarifies a fundamental notion in quantum
mechanics.

\end{abstract}

%-----------------------------------------------------------------------------

\newpage
\section{Introduction}
\label{Intro}

The topic of this paper concerns a simple formula, rarely mentioned in the
literature, which can serve as a helpful tool in quantum mechanics. It has
been shown \cite {AV} that for any Hermitian operator $A$ and any quantum
state $|\psi\rangle$, the following formula is valid:
\begin{equation}
A |\psi\rangle = \langle A \rangle |\psi\rangle + \Delta A |\psi_{\perp}
\rangle , \label{formula}
\end{equation}
where $|\psi\rangle$, $|\psi_{\perp}\rangle$ are normalized vectors,
$\langle\psi_{\perp}|\psi\rangle = 0$, $\langle A \rangle \equiv
\langle\psi|A|\psi\rangle$, and $\Delta A \equiv \sqrt{\langle A^2 \rangle -
\langle A \rangle^2}$.
The proof is as follows: It is always possible to decompose $A |\psi\rangle =
\alpha |\psi\rangle + \beta |\psi_{\perp}\rangle$ with $\beta \geq 0$.
Then $\langle\psi| A |\psi\rangle = \langle\psi| \, (\alpha |\psi\rangle +
\beta |\psi_{\perp}\rangle)$ yields $\alpha = \langle A \rangle$, and
$\langle\psi| A^{\dagger} A |\psi\rangle = (\alpha^{\star} \langle\psi| +
\beta^{\star} \langle\psi_{\perp}|) \, (\alpha |\psi\rangle + \beta
|\psi_{\perp}\rangle)$ so that $\beta = \Delta A$.

In ref. \cite {AV}  the formula has been applied to a composite system
consisting of a large number of parts in a product state. It was proven that
such a product state is essentially an eigenstate of an operator defined as
an ``average'' of variables corresponding to these parts. The formula has also
been used in a simple derivation of the minimal time for the evolution of a
quantum system to an orthogonal state \cite {V}. Our aim here is to show new
applications of this formula. In section \ref{HighDelta} an immediate result,
related to the uncertainty of an operator in two orthogonal states, is
obtained. In section \ref{Paradox} we present an apparent paradox which arises
when the formula is used in a naive way. In section \ref{Uncertain} we use it
to derive in a simple way the Heisenberg uncertainty principle and other
related inequalities.

%-----------------------------------------------------------------------------

\section{A Maximal Uncertainty State is Not Unique}
\label{HighDelta}

Let us rewrite our basic formula in the form
\begin{equation}
A \, |\psi\rangle = \langle A \rangle_{\psi} \, |\psi\rangle +
\Delta A_{\, \psi} \, |\psi_{\perp} \rangle . \label{formula-A-psi}
\end{equation}
Then, the scalar product of $|\psi_{\perp} \rangle$ and $A \, |\psi\rangle$ is
\begin{equation}
\langle \psi_{\perp}| A |\psi\rangle = \Delta A_{\, \psi} . \label{calc20}
\end{equation}
For $A \, |\psi_{\perp} \rangle$ the formula gives
\begin{equation}
A \, |\psi_{\perp}\rangle = \langle A \rangle_{\psi_{\perp}} \,
|\psi_{\perp}\rangle +
\Delta A_{\, \psi_{\perp}} \, |\psi_{\perp \perp} \rangle .
\label{formula-A-psi_perp}
\end{equation}
where $\langle\psi_{\perp \perp}|\psi_{\perp}\rangle = 0$.
Substituting eq.(\ref{formula-A-psi_perp}) in eq.(\ref{calc20}) yields
\begin{equation}
\Delta A_{\, \psi} = \Delta A_{\, \psi_{\perp}} \,
\langle \psi|\psi_{\perp \perp} \rangle . \label{calc25}
\end{equation}
We see that in the case of a two-dimensional Hilbert space where only two
mutually orthogonal states exist, $\Delta A_{\, \psi} =
\Delta A_{\, \psi_{\perp}}$. Since $|\langle \psi|\psi_{\perp \perp} \rangle|
\leq 1$, eq.(\ref{calc25}) leads to
\begin{equation}
\Delta A_{\, \psi_{\perp}} \geq \Delta A_{\, \psi} . \label{calc30}
\end{equation}
Thus, we have proved the following theorem:
\begin{description}
\item \quad \quad For any Hermitian operator $A$ and any given state
$|\psi\rangle$ there exists a state $|\psi_{\perp} \rangle$ orthogonal to
$|\psi\rangle$, such that $\Delta A_{\, \psi_{\perp}} \geq
\Delta A_{\, \psi}$.
\end{description}
This implies that a state corresponding to a maximal uncertainty of any given
observable cannot be unique.

%-----------------------------------------------------------------------------

\section{An Apparent Paradox}
\label{Paradox}

Let us consider a system described by a two-dimensional Hilbert space, such
as a spin-$\frac{1}{2}$ particle. Then, for an Hermitian operator $B$,
different than $A$, a relation similar to eq.(\ref{formula}) holds:
\begin{equation}
B |\psi\rangle = \langle B \rangle |\psi\rangle + \Delta B |\psi_{\perp}
\rangle , \label{formula-B}
\end{equation}
where the vector $|\psi_{\perp}\rangle$ is the same as in eq.(\ref{formula})
(since it is the only vector orthogonal to $|\psi\rangle$). Note that the
quantities $\langle A \rangle$, $\langle B \rangle$, $\Delta A$ and $\Delta B$
are all real numbers. Multiplying the Hermitian conjugate of
eq.(\ref{formula-B}) by eq.(\ref{formula}), and using the fact that
$B = B^{\dagger}$, we obtain
\begin{equation}
\langle\psi| B A |\psi\rangle =
(\langle B \rangle \langle\psi| + \Delta B \langle\psi_{\perp}|) \:
(\langle A \rangle |\psi\rangle + \Delta A |\psi_{\perp}\rangle) =
\langle B \rangle \langle A \rangle + \Delta B \, \Delta A . \label{<BA>}
\end{equation}
Also, multiplying the Hermitian conjugate of eq.(\ref{formula}) by
eq.(\ref{formula-B}) yields
\begin{equation}
\langle\psi| A B |\psi\rangle =
\langle A \rangle \langle B \rangle + \Delta A \, \Delta B . \label{<AB>}
\end{equation}
Thus, subtracting eq.(\ref{<BA>}) from eq.(\ref{<AB>}) leads to
\begin{equation}
\langle [A,B] \rangle = 0 . \label{<A,B>}
\end{equation}
Equation (\ref{<A,B>}) states that the expectation value of the commutator of
two {\it arbitrary} operators is always zero, irrespective of the wavefunction
which describes the system.

This ``remarkable'' result can be simply tested using an example of a
spin-$\frac{1}{2}$ particle, where $A = \sigma_x$, $B = \sigma_y$, and
$|\psi\rangle = |\uparrow_z\rangle$. The calculation yields
\begin{equation}
\langle [A,B] \rangle = \langle \uparrow_z| [\sigma_x,\sigma_y] |\uparrow_z
\rangle = \langle \uparrow_z| 2 \, i \, \sigma_z |\uparrow_z \rangle = 2 \, i
, \label{2i}
\end{equation}
in contradiction to eq.(\ref{<A,B>}).

A paradox? Not really. Obviously, something is wrong in the derivation of
eq.(\ref{<A,B>}). Each of the two basic equations (\ref{formula}) and
(\ref{formula-B}) is separately correct, however, they are not correct when
used together. They both use the same vector $|\psi_{\perp}\rangle$ orthogonal
to $|\psi\rangle$, but even in a two-dimensional Hilbert space the orthogonal
vector is uniquely defined only up to a phase. If we use the vector
$|\psi_{\perp}\rangle$ as it is defined by eq.(\ref{formula}), then we should
rewrite eq.(\ref{formula-B}) as follows:
\begin{equation}
B |\psi\rangle = \langle B \rangle |\psi\rangle + \Delta B \, e^{i \varphi} \,
|\psi_{\perp} \rangle , \label{formula-B-correct}
\end{equation}
where $\varphi$ is real.
Equation (\ref{<A,B>}) is now replaced by the correct expression
\begin{equation}
\langle [A,B] \rangle = \Delta A \, \Delta B \, (e^{i \varphi} -
e^{-i \varphi}) = 2 \, i \, \Delta A \, \Delta B \, \sin{\varphi} .
\label{<A,B>-correct}
\end{equation}
Only if $\varphi = 2 \pi n$ where $n = 0, \pm 1, \pm 2, ...$, the expectation
value of $[A,B]$ is equal to zero -- not always, as concluded above.

Let us reexamine the previous example with $A = \sigma_x$, $B = \sigma_y$,
and $|\psi\rangle = |\uparrow_z\rangle$.
\begin{eqnarray}
A |\psi\rangle & = & \sigma_x |\uparrow_z\rangle = |\downarrow_z\rangle ,
\label{calc40} \\
B |\psi\rangle & = & \sigma_y |\uparrow_z\rangle = i |\downarrow_z\rangle .
\label{calc45}
\end{eqnarray}
Comparing these equations with eqs.(\ref{formula}) and
(\ref{formula-B-correct}) we obtain $\Delta \sigma_x = \Delta \sigma_y = 1$
and $e^{i \varphi} = i$ (that is $\sin{\varphi} = 1$). Introducing these
values in eq.(\ref{<A,B>-correct}) we find
\begin{equation}
\langle \uparrow_z| [\sigma_x,\sigma_y] |\uparrow_z \rangle = 2 \, i ,
\label{calc50}
\end{equation}
in perfect agreement with eq.(\ref{2i}). Summing up, the phase in quantum
mechanics is too important to be neglected.

%-----------------------------------------------------------------------------

\section{The Heisenberg Uncertainty Principle}
\label{Uncertain}

Our understanding of the preceding apparent paradox has provided us with
useful algebraic tools to be implemented in this section. We present here a
simple method, based on the formula (eq.(\ref{formula})), to obtain the
Heisenberg uncertainty principle. Consider two Hermitian operators, $A$ and
$B$, in an arbitrary Hilbert space. Then, the following equations hold:
\begin{eqnarray}
A |\psi\rangle & = & \langle A \rangle |\psi\rangle + \Delta A |\psi_{\perp A}
\rangle , \label{formula-A-new} \\
B |\psi\rangle & = & \langle B \rangle |\psi\rangle + \Delta B |\psi_{\perp B}
\rangle , \label{formula-B-new}
\end{eqnarray}
where $\langle\psi_{\perp A}|\psi\rangle = 0$ and
$\langle\psi_{\perp B}|\psi\rangle = 0$.
Following the same procedure as in section \ref{Paradox}, we find that
\begin{equation}
\langle B A \rangle = \langle B \rangle \langle A \rangle + \Delta B \,
\Delta A \, \langle\psi_{\perp B}|\psi_{\perp A}\rangle , \label{<BA>-new}
\end{equation}
and
\begin{equation}
\langle A B \rangle = \langle A \rangle \langle B \rangle + \Delta A \,
\Delta B \, \langle\psi_{\perp A}|\psi_{\perp B}\rangle . \label{<AB>-new}
\end{equation}
Subtracting eq.(\ref{<BA>-new}) from eq.(\ref{<AB>-new}) yields
\begin{eqnarray}
\langle [A,B] \rangle & = &
\Delta A \, \Delta B \, (\langle\psi_{\perp A}|\psi_{\perp B}\rangle -
\langle\psi_{\perp B}|\psi_{\perp A}\rangle) \nonumber \\ & = &
2 \, i \, \Delta A \, \Delta B \: {\cal I}m \langle\psi_{\perp A}|\psi_{\perp
B}\rangle . \label{<A,B>-new}
\end{eqnarray}
Then, taking the absolute value of eq.(\ref{<A,B>-new}) we find
\begin{equation}
\Delta A \, \Delta B \: |{\cal I}m \langle\psi_{\perp A}|\psi_{\perp B}\rangle|
= \frac{1}{2} \: |\langle [A,B] \rangle| . \label{calc5}
\end{equation}
The vectors are normalized, therefore
$|{\cal I}m \langle\psi_{\perp A}|\psi_{\perp B}\rangle| \leq 1$, so that we
end with
\begin{equation}
\Delta A \, \Delta B \geq \frac{1}{2} \: |\langle [A,B] \rangle| ,
\label{Heisen}
\end{equation}
which is the standard form of the uncertainty principle.

Another interesting inequality can be obtained by calculating the
anti-commutator of $A$ and $B$. We add eq.(\ref{<BA>-new}) to
eq.(\ref{<AB>-new}) and get
\begin{eqnarray}
\langle \{A,B\} \rangle & = &
2 \, \langle A \rangle \langle B \rangle +
\Delta A \, \Delta B \: (\langle\psi_{\perp A}|\psi_{\perp B}\rangle +
\langle\psi_{\perp B}|\psi_{\perp A}\rangle) \nonumber \\ & = &
2 \, \langle A \rangle \langle B \rangle +
2 \, \Delta A \, \Delta B \: {\cal R}e \langle\psi_{\perp A}|\psi_{\perp
B}\rangle , \label{<[[A,B]]>}
\end{eqnarray}
where $\{A,B\} = AB + BA$.
Rearranging eq.(\ref{<[[A,B]]>}) and taking the absolute values of both
sides, we find
\begin{equation}
\Delta A \, \Delta B \: |{\cal R}e \langle\psi_{\perp A}|\psi_{\perp B}\rangle|
= \left| \frac{1}{2} \, \langle \{A,B\} \rangle - \langle A \rangle \langle B
\rangle \right|.
\end{equation}
Since $|{\cal R}e \langle\psi_{\perp A}|\psi_{\perp B}\rangle| \leq 1$, it
follows that
\begin{equation}
\Delta A \, \Delta B \geq \left| \frac{1}{2} \, \langle \{A,B\} \rangle -
\langle A \rangle \langle B \rangle \right| . \label{Ineq}
\end{equation}
This inequality is a by-product of a conventional derivation of the
uncertainty principle which is based on the Cauchy-Schwarz inequality
\cite{Vider}. The physical significance of eq.(\ref{Ineq}) is that it provides
an estimate for the correlations developed in time between $A$ and $B$. For
instance, it manifests the correlation between $x$ and $p$ for the case of a
free particle evolving in time \cite{Bohm}.

We can also obtain a more accurate expression for $\Delta A \, \Delta B$.
Adding eq.(\ref{<A,B>-new}) to eq.(\ref{<[[A,B]]>}) we find
\begin{eqnarray}
\langle [A,B] \rangle + \langle \{A,B\} \rangle & = &
2 \, i \, \Delta A \, \Delta B \: {\cal I}m \langle\psi_{\perp A}|\psi_{\perp
B}\rangle \nonumber \\ & \phantom{.} & + \;
2 \, \Delta A \, \Delta B \: {\cal R}e \langle\psi_{\perp A}|\psi_{\perp
B}\rangle +
2 \, \langle A \rangle \langle B \rangle ,
\end{eqnarray}
and consequently,
\begin{equation}
\Delta A \, \Delta B \: \langle\psi_{\perp A}|\psi_{\perp B}\rangle =
\frac{1}{2} \, \langle [A,B] \rangle + \frac{1}{2} \, \langle \{A,B\} \rangle
- \langle A \rangle \langle B \rangle .
\end{equation}
Taking the norm of both sides we find
\begin{equation}
\Delta A \, \Delta B \geq \left| \frac{1}{2} \, \langle [A,B] \rangle +
\frac{1}{2} \, \langle \{A,B\} \rangle -
\langle A \rangle \langle B \rangle \right| . \label{calc10}
\end{equation}
Since $[A,B] = i \, C$ and $\{A,B\} = D$, where $C$ and $D$ are Hermitian
operators, and since the expectation value of an Hermitian operator is a real
number, it follows that
\begin{equation}
\Delta A \, \Delta B \geq \left[
\left( \frac{1}{2} \, \langle \{A,B\} \rangle -
\langle A \rangle \langle B \rangle \right)^2 +
\frac{1}{4} \, |\langle [A,B] \rangle|^{\:2}
\right]^{\frac{1}{2}} . \label{ComplexIneq}
\end{equation}
This result combines the two previously found bounds, i.e. eqs.(\ref{Heisen})
and (\ref{Ineq}).

%-----------------------------------------------------------------------------

%-----------------------------------------------------------------------------


\bigskip \bigskip
\begin{thebibliography}{99}

\bibitem{AV}  Y. Aharonov and L. Vaidman, ``Properties of a Quantum System
During the Time Interval Between Two Measurements'', Phys. Rev. A {\bf 41},
11 (1990).

\bibitem{V}  L. Vaidman, ``Minimal Time for the Evolution to an Orthogonal
State'', Am. J. Phys. {\bf 60}, 182 (1992).

\bibitem{Vider}  See, for instance, S. Wieder, {\it The Foundations of Quantum
Theory} (Academic Press, New York, 1973), pp. 64-65.

\bibitem{Bohm} D. Bohm, {\it Quantum Theory} (Prentice-Hall, Englewood Cliffs,
NJ, 1951), pp. 203-207.


\end{thebibliography}
\end{document}